\documentclass{article}

\usepackage{arxiv}

\usepackage[utf8]{inputenc} %
\usepackage[T1]{fontenc}    %
\usepackage{hyperref}       %
\usepackage{url}            %
\usepackage{booktabs}       %
\usepackage{amsfonts}       %
\usepackage{nicefrac}       %
\usepackage{microtype}      %
\usepackage{lipsum}
\usepackage{graphicx}
\graphicspath{ {./images/} }
\usepackage{xcolor}

\title{From Statistics to Individuals: An Exploration of Zoomable Empathic Visualizations}

\author{
 Edwige Chauvergne \\
  LyRIDS, ECE Engineering School, OMNES
Education\\
  France \\
  \texttt{echauvergne@ece.fr} \\
   \And
  Arnaud Prouzeau \\
  Inria, CNRS\\
  France \\
  \texttt{arnaud.prouzeau@inria.fr} \\
  \And
  Martin Hachet \thanks{Bivwac and LaBRI labs}\\
  Inria, CNRS \\
  France \\
  \texttt{martin.hachet@inria.fr} \\
  \And
  Pierre Dragicevic\\
  Inria, CNRS\\
  France \\
  \texttt{pierre.dragicevic@inria.fr} \\
}

\newcommand{\excerpt}[1]{\textcolor[rgb]{0.4,0.4,0.4}{``#1''}}

\begin{document}

\maketitle

\begin{figure*}[h]
 \includegraphics[width=0.9\linewidth]{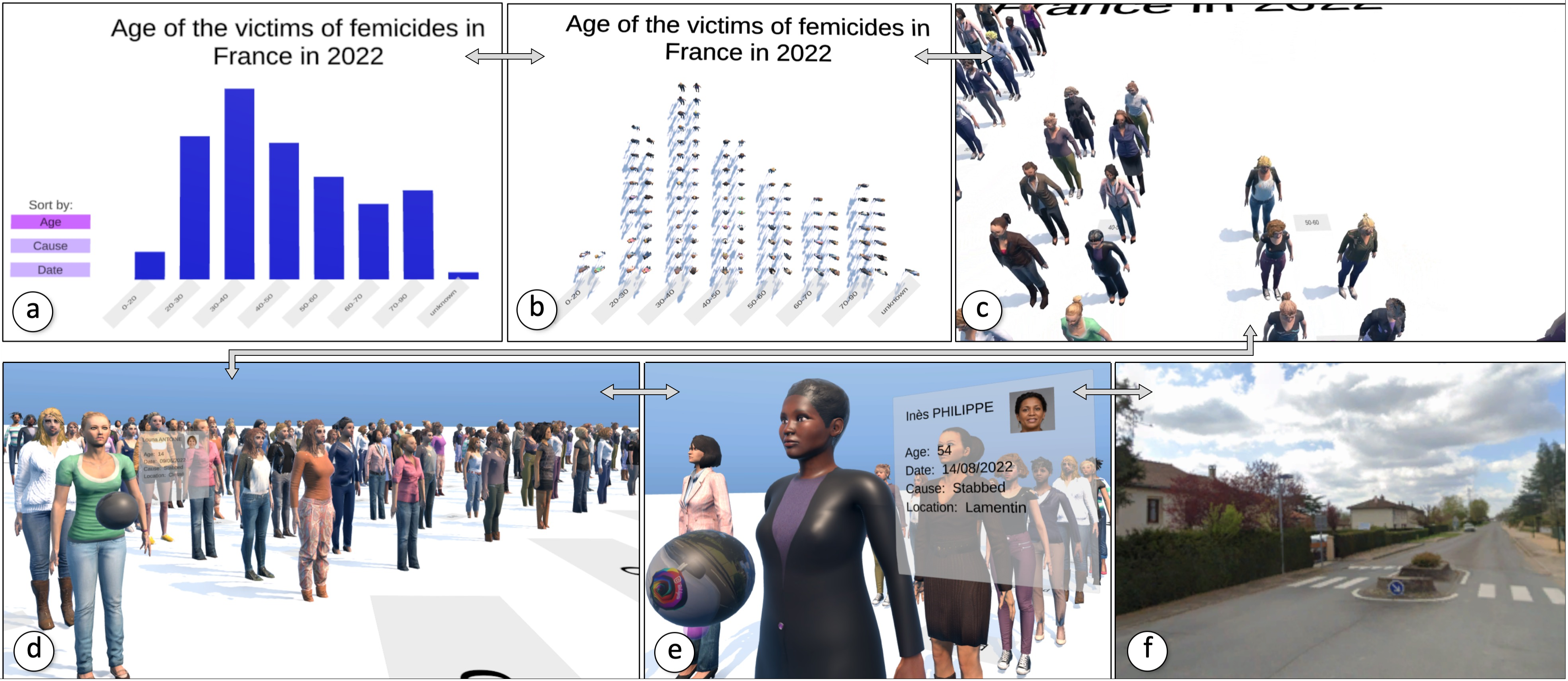}
 \centering

  \caption{Screenshots of a zoomable empathetic visualization (ZEV) showing data about femicides at different zoom levels. \textit{(a)} the initial view is a classic bar chart. \textit{(b)} the bar chart turns into a unit chart where each victim is represented by a square~; squares then morph into 3D avatars viewed from the top. \textit{(c)} the user ends up ``landing'' on the visualization, \textit{(d)} finding themselves inside the crowd of femicide victims. \textit{(e)} when the user approaches a particular victim, a box shows information about the victim. The user can grab the sphere next to the victim, in order to \textit{(f)} immerse themselves into a 360° view of the neighborhood where the femicide took place. Names and photos have been changed.}
\label{fig:teaser}
\end{figure*}

\begin{abstract}
Data visualization is a powerful tool for conveying statistical information, but when representing populations, it tends to hide individuals. We introduce Zoomable Empathic Visualizations (ZEVs), interactive experiences allowing users to smoothly navigate between abstract statistical visualizations and more qualitative, relatable representations focused on individuals. 
We present three use cases of ZEVs and report on a qualitative user study that highlights opportunities for deeper understanding and emotional engagement, while pointing to areas for improvement and further refinement. In summary, ZEVs point toward new approaches for revealing the individuals behind the data.  
\end{abstract}

\keywords{Virtual reality \and Humanitarian data \and Storytelling \and Semantic zoom \and Empathic visualization}

\section{Introduction}

Many human tragedies are unfolding worldwide, and effectively communicating about these tragedies can be a first step towards improvement. Behind many tragedies are numbers, and knowing those numbers can be important to build an understanding. For example, in 1924, over 13\% of children in France died before reaching the age of five, whereas by 2023, this figure had dropped to 0.4\% \cite{FrenchDatasetInfant}---a dramatic drop that can be effectively conveyed by a simple bar chart or a line chart \cite{owid-child-mortality-global-overview}. However, such a chart can inadvertently obscure the suffering of the many children and parents who lived this tragedy in 2023. More generally, charts and data visualizations can greatly help people understand both simple numbers and complex data \cite{munzner2014visualization}, but due to their abstract nature, they can obscure the suffering of victims. As Jacob Harris \cite{harris2015connecting} mentioned, \textit{``from a distance, it’s easy to forget the dots are people''}. Dragga and Voss \cite{dragga2001cruel} go as far as stating that conveying human tragedies using abstract visualizations is unethical.

To promote a deeper understanding of human tragedies and bring out the common humanity between the audience and the victims, communicators and designers often rely on stories and narratives \cite{Hiltunen2019Documentary}. For example, immersive documentaries offer realistic reconstructions of human tragedies, and sometimes let users take the perspective of protagonists \cite{rodriguez2020use}, such as a young teenager in a refugee camp \cite{united2016clouds}. Narratives, immersive or otherwise, help understand human experiences on a visceral level and are generally more effective at making people care than numbers. For example, people are more likely to donate after viewing a photo of a victim than being told the number of victims \cite{Lee2016Identifiable}. However, stories convey a limited quantitative understanding of human tragedies. They provide an incomplete---potentially distorted \cite{van2022news}---picture of reality, which can impair understanding and fair decision making, for example, when deciding how to allocate helping resources \cite{caviola2021psychology,verma2023designing}.

We therefore have on one side \textit{quantitative information} that provides a crucial type of information (e.g., the extent of a humanitarian crisis), and on the other side \textit{qualitative information} that provides another type of crucial information (e.g., an intuitive understanding of the personal experience of victims). Recently, we have seen research trying to combine both worlds. In particular, there has been a surge of interest in \textit{narrative visualization}, where data visualizations are combined with conventional storytelling techniques and media such as newspaper articles and videos \cite{segel2010narrative,tong2018storytelling}. There have also been explorations of data visualizations that emphasize individuals \cite{morais2020showing,morais2021can} -- a research area  called \textit{anthropographics}. Although this area is promising, it is still nascent, without clear empirical evidence for a benefit of new designs so far \cite{morais2021can}. In both narrative visualizations and anthropographics, data and stories are intertwined, but they are rarely seamlessly blended: many existing approaches maintain a divide between the world of data and the world of stories. In addition, most current narrative visualizations and anthropographics are data-oriented. In particular, we are not aware of any narrative data visualization that conveys rich information about individuals, as news articles or documentaries do. To summarize, research has not fully explored interactive systems that try to break the divide between numbers and stories, where people can understand the two facets of human tragedies in a holistic manner.

As a step towards this goal, we introduce \textit{Zoomable Empathic Visualizations} (ZEVs) (see \autoref{fig:teaser} for an example). ZEVs are interactive experiences that let users smoothly navigate back-and-forth between statistical data visualizations and representations that are more concrete, qualitative, individualized, emotional, and immersive. ZEVs are based on the \textit{semantic zooming} user interface technique \cite{perlin1993pad}, where users navigate in visual information by zooming, and where different types of visual information are shown at different zoom levels. In a typical ZEV, the user starts with an abstract graph, and by zooming in, progressively moves towards views that give richer  information about individuals and are increasingly immersive, up to immersive experiences allowing the user to embody individuals. We expect this new type of visualization to be both educational and emotionally impactful, and raise awareness by addressing both the user's rationality and emotions.

In this paper, we make the following contributions: 1) we introduce the concept of ZEV, a type of interactive visualization that aims at bridging the gap between abstract aggregated visualizations and rich visceral experiences; 2) based on a proof-of-concept prototype, we demonstrate three use cases for ZEVs whose purpose is to communicate about victims of femicides, bicycle accidents, and farmed-animal conditions; 3) we report a qualitative study that highlights some of the strengths and drawbacks of ZEVs.%

\section{Background}

Here we provide a broad survey of related work, covering work in visualization, storytelling, and virtual reality.

\subsection{Conveying Tragedies: Data vs. Stories}

Sharing data is one way of informing people about humanitarian crises. Humanitarian data has been defined as 
\textit{``data about the context in which a humanitarian crisis is occurring (e.g. baseline/development data, damage assessments, geospatial data); about the people affected by the crisis and their needs; or about the response by organizations and people seeking to help those who need assistance''} \cite{humdata2024}. Visualizing such data can help decision makers and general audiences understand the quantitative aspects of a humanitarian crisis.

For general audiences, a more common communication strategy is to share stories about victims. Although stories may give little information regarding the number of victims affected by an event, they can generate a visceral and emotional experience that is often more impactful than plain numbers \cite{Lee2016Identifiable, Mencarini2025Stories}. Fictional stories are used as well. For example, the short movie ``Most Shocking Second-a-Day Video'' \cite{stirlig2014} draws a parallel with the Syrian war by depicting a young Londonian girl going through a hypothetical war.

In the recent years, researchers have explored how to visualize data about people in a way that helps the audience relate to the individuals \cite{Boy2017Showing,morais2020showing,morais2021can}. This area of research has been called \textit{anthropographics}, which Morais et al. \cite{morais2020showing} broadly define as \textit{``visualizations that represent data about people in a way that is intended to promote prosocial feelings (e.g., compassion or empathy) or prosocial behavior (e.g., donating or helping)''}. 
A common strategy is to ``de-aggregate'' visualizations so that each individual is represented by a separate symbol \cite{morais2020showing}. Such visualizations are also called \textit{unit charts} \cite{park2017atom}.

Several studies were conducted to evaluate anthropographic design strategies, with mostly inconclusive results \cite{Boy2017Showing,morais2021can, liem2020structure}. However, studies merely used visual symbols to represent individuals, without conveying information about those individuals. Therefore, one new direction has been to convey actual information about the victims \cite{morais2021can}. Dhawka et al. \cite{Dhawka2023WeAreTheData} similarly argued for representing sociodemographic diversity in visualizations about people. Although there is presently no strong evidence that conveying basic information about individuals can help the audience relate \cite{morais2021can}, there has been no study so far on conveying rich information, a concept we explore here with Zoomable Empathic Visualizations (ZEVs).

Raising empathy for a large number of victims is notoriously challenging. A number of studies have found that empathy decreases as the number of victims increases, a phenomenon known as \textit{compassion fade} \cite{slovic2007if, vastfjall2014compassion} or \textit{scope neglect} \cite{caviola2021psychology}. 
This phenomenon is thoroughly discussed and described in Paul Slovic's work \cite{slovic2015numbers}.
A related psychological phenomenon is the \textit{identifiable victim effect}, whereby people donate more to victims whom they can identify (e.g., with a photo or a name) than to anonymous victims \cite{small2003helping}. Both psychological effects can lead people to direct their emotional energy and resources to small causes, ignoring issues causing large amounts of suffering. Although there is no simple solution to this problem, it may be partly alleviated by conveying both the scope of a crisis (with data visualizations) and personal details about the victims themselves (with qualitative information displays and stories). However, compassion fade can partly cancel the benefits brought by conveying information about victims, as people donate more to a single identified victim than to a group of identified victims \cite{kogut2005identified}.

When sharing information about individual victims, one faces a trade-off between the risks of invading privacy and the ethical issues raised by the anonymity of victims. A study on the representation of refugees in Australian media \cite{bleiker2013visual} argued that as migrants are shown in crowded boats rather than individually, they are dehumanized and more easily perceived as a threat. Some data journalists similarly advocate for showing the people behind the data \cite{harris2015connecting, slobin2014WhatIf}. In their paper \textit{``Cruel pies: The inhumanity of technical illustrations''}, Dragga and Voss \cite{dragga2001cruel} argue that conventional visualizations are inhumane because they look the same whether they refer to human victims or inanimate objects.

\subsection{Virtual Reality as a Medium}

Virtual reality (VR) has been explored as an alternative information medium, both in data visualization and in traditional storytelling.

Recently, \textit{immersive analytics} has emerged as a new research area in visualization,  whose goal is to study how can immersive displays facilitate the analysis of complex datasets \cite{ens2021grand}. More recently, researchers have explored the use of familiar objects in VR to represent numbers and quantities in a visceral way \cite{lee2020data,assor2024augmented}.

Despite the work done in this area, most virtual-reality visualization systems proposed so far do not focus on conveying data about human tragedies. One notable exception is the immersive visualization of mass shooting victims by Ivanov et al. \cite{Ivanov2019AWalk}, a concept that ZEVs extend and generalize. Specifically, we extend this work by investigating the transition from traditional charts to individual-focused immersive experience, and by conducting an exploratory study to better understand the potential and limitations of such a visualization.

Over the past decade, VR has been often used to tell stories, including about human tragedies \cite{bucher2017storytelling}. The concept of immersive journalism was introduced about 15 years ago and defined as \textit{``the production of news in a form in which people can gain first-person experiences of the events or situation described in news stories''} \cite{Pena2010immersive}. 
These VR stories can take various formats. For example, the United Nations have produced several 360° videos to raise awareness of children's living conditions in Syrian refugee camps \cite{united2016clouds} and in Haiti \cite{united2019Haiti}. In \textit{Titanic VR} \cite{titanicVR2014}, the user embodies a passenger on a lifeboat while watching the Titanic sink. In some virtual experiences, the user can move and interact with the virtual content \cite{centerEResearch2022cut} -- they become an agent \cite{arcagni2021vr}. VR can even be used to give people the illusion that they inhabit other bodies \cite{kilteni2012sense}.

So far there is little work combining immersive data visualization with immersive stories. Some recent work has started to extend the concept of narrative visualization (i.e., data visualizations with narrative structures \cite{segel2010narrative}) to immersive displays. For example, \textit{Flow Immersive} \cite{dibenigno2021flow} is an immersive and interactive data visualization tool that integrates storytelling components into the visualizations to make the data more accessible to a large audience. However, there is very little work looking at how such techniques could be used to communicate about human tragedies specifically, with the exception of a position paper that introduces the concept of \textit{immersive humanitarian visualization} and speculates about its possible benefits, but without proposing a design or an implementation \cite{dragicevic2022towards}.

\subsection{Zoomable UIs and Semantic Zooming}

The term \textit{zoomable user interface} (\textit{ZUI}) was coined in 1993 \cite{perlin1993pad,bederson1994pad++} and later defined as \textit{``systems that support the multi-scale and spatial organisation of and magnification-based navigation among multiple documents or visual objects''} \cite{bederson2011promise}. Most ZUIs employ a technique called \textit{semantic zooming}, where the appearance of visual elements differs depending on the zoom level \cite{perlin1993pad}. In data visualization, semantic zooming can be used to drill in and out a dataset \cite{weaver2004building}. Semantic zooming has been the subject of multiple studies in HCI \cite{dunsmuir2009selective, conti2005visual, garcia2011semantic} and is nowadays widely used in map applications. ZEVs use semantic zooming as their main navigation paradigm for letting users explore humanitarian data at different levels of detail. 

\section{Examples of Zoomable Empathic Visualizations}
\label{sec: use cases}

In this section, we illustrate Zoomable Empathic Visualizations (ZEVs) with three use cases, developed using a proof-of-concept VR prototype. A video is available in our \href{https://osf.io/awqgc/}{OSF repository}. 

\subsection{Use Case 1: Femicides}
\label{sec: use case feminicides}
In this first use case, a ZEV is used to visualize data about suspected cases of femicides having taken place in France in 2022, and published by the collective \textit{``Féminicides par compagnons ou ex''} \cite{Femicides}. This dataset contains data on the individual victims such as their name, picture, age and context of death (the victims' names and photos were anonymized on the \href{https://osf.io/awqgc/}{OSF repository}, but not in the study reported in \autoref{sec:study}). The prototype was already mentioned in the introduction and illustrated in \autoref{fig:teaser}.

As shown in \autoref{fig:teaser}-a, the initial visualization is a conventional 2D bar chart showing the number of cases per victim age. The user sees the chart in front of them in their VR headset, as if it was on a distant wall. On the left, buttons allow the user to sort the cases differently, i.e., by cause or time of death. At this stage, the user can already get high-level quantitative insights: for example, that femicides happened at all ages but were more prevalent at the ages of 30 and 40; that they took place at any time of the year, and involved various causes of deaths (15 causes are listed), but that the two major causes were stabbing and firearm shooting.

\begin{figure*} 
    \centering
    \includegraphics[width=0.9\textwidth, alt={
    Four figures: fig A and B depict the bike accident use case and fig C and D depict the hen welfare use case.
    Fig. a: Second step of the bicycle accident use cases. The bar charts is divided into unit cubes. 
    Fig. b: Close visualization of an avatar representing a female victim. The panel next to her indicates that it happened on January 10th 2018, that she was 40 year-old, and it happened with a covered weather. The city where is happened is anonimized
    Fig. c: Vertical visualization of the avatars. Each avatar is 1-meter square and is composed of as many hens avatars as the number of hens per square meter allowed for this type of farm.
    Fig. d: Close visualization of hens in ground farming conditions. On the panel, it is written "Inside density: 9 hen/m2  outside density: No access  Flock size: No limit (up to 20000 and more)"}
    ]
    {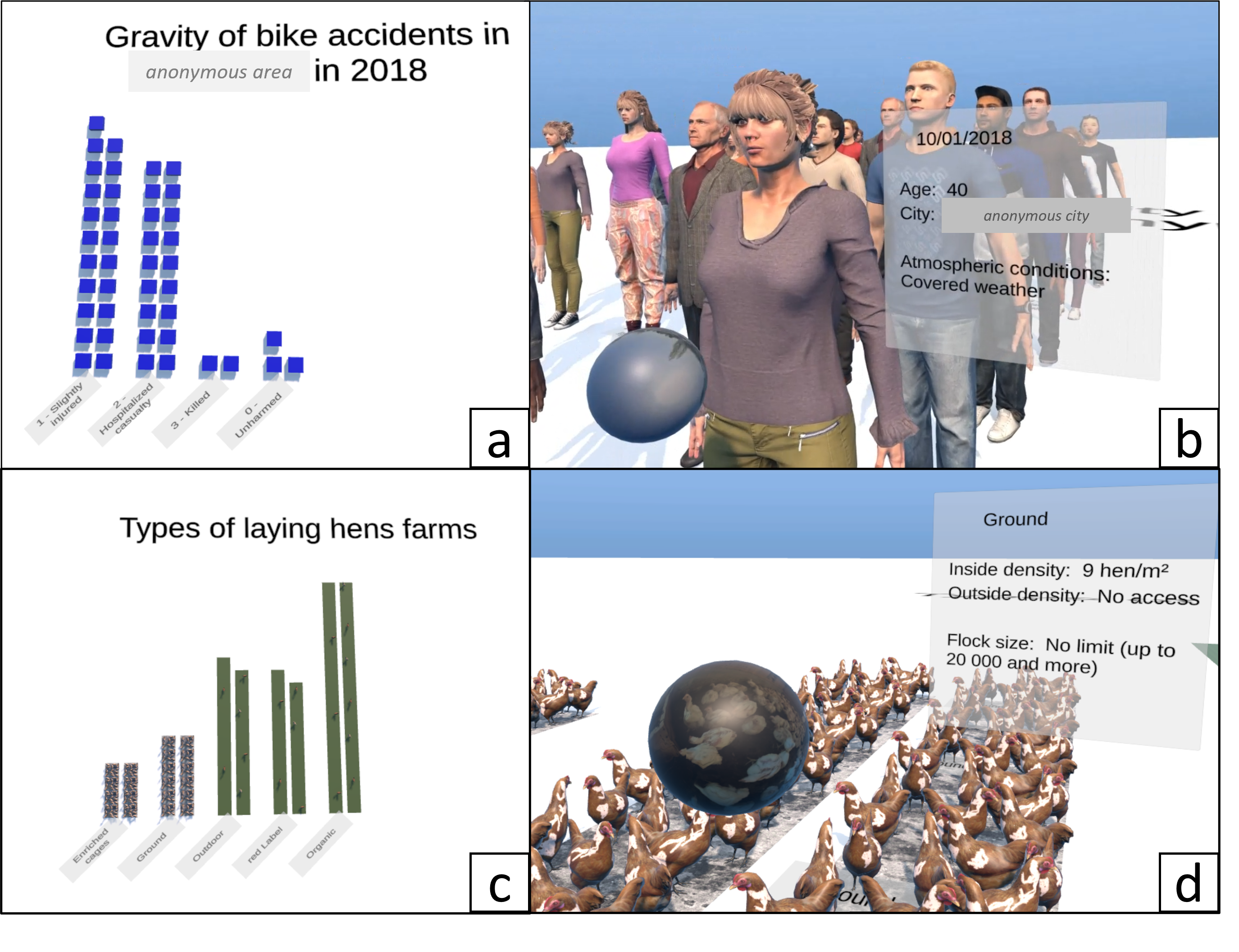}
    \caption{Screenshots from the ZEV uses cases showing bicycle accidents data \emph{(a and b)} and hen welfare data \emph{(c and d)}. 
    \emph{(a)} unit chart showing cases sorted by their gravity (slightly injured, hospitalized, killed, unharmed); 
    \emph{(b)} avatar and infobox showing information about a victim and the accident in which they were involved.
    \emph{(c)} Top view of the unit chart with hen avatars, showing hen density for different categories of farms (enriched cages, ground, outdoor, red label, organic);
    \emph{(d)} by walking among hen avatars, the user can get a better sense of the hen density, read information about the farming conditions, and immerse themselves into a 360° view of a typical farm.
    }
    \label{fig:bike_hens_examples}
\end{figure*}

The user can pan and zoom the 2D chart using the two VR controllers. When the user zooms in, the bar chart animates into a unit chart \cite{park2017atom}, where each square stands for a single person (not shown in \autoref{fig:teaser}, but \autoref{fig:bike_hens_examples}-a shows a similar representation). At this stage, the user can see the number of individuals making up each bar of the bar chart. Zooming in further turns squares into 3D avatars (\autoref{fig:teaser}-b). At this stage, avatars are far away and their details cannot be seen.

If the user keeps zooming in, the visualization tips over and becomes horizontal, and the user lands on it (\autoref{fig:teaser}-c and d). The user now stands among life-sized human avatars, and can freely walk between them. In this example, the avatars were manually selected to resemble the victims’ photographs when available. This reinforces the message that each victim was a real individual, while also highlighting the diversity of their visible demographic characteristics—such as age, skin color, and other physical traits. When the user approaches a victim, a box shows extra information about the victim and the circumstances of their death (right side of \autoref{fig:teaser}-e).

When the user approaches a victim, a sphere also appears next to them (left side of \autoref{fig:teaser}-e). If the user grabs the sphere and brings it close to their face\footnote{Both this interaction and the landing interaction mentioned previously were inspired by the Google Earth VR app \cite{GoogleEarthVR}.}, they get teleported into a 360° view of the place or neighborhood where the femicide took place (\autoref{fig:teaser}-f). This allows the user to realize that femicides took place in various types of environments: cities, countryside, suburbs, and so on.

At any point in time, the user can use the VR controllers to teleport themselves to any location in the immersive visualization or zoom back to a top view.

\subsection{Use Case 2: Bicycle Accidents \label{sec: use case bike}}
In this second use case, a ZEV is used to visualize data about bicycle accidents recorded by law enforcement, which took place in 2018 in a specific French region. This data was extracted from a governmental dataset covering bike accidents in France from January 2005 to December 2018 \cite{BikeAccidents}.

The different visualizations are similar to the previous example in their design, and the interactions are the same. As before, the application starts with a simple bar chart, this time showing bicycle accidents sorted by severity (from unharmed to killed). The user can alternatively sort by lighting condition, victims' age, and date. At this stage, the user can see for example that accidents affected all age groups, and the vast majority occurred in broad daylight.

As before, zooming in turns the bar chart into a unit chart (\autoref{fig:bike_hens_examples}-a), which become human avatars at higher zoom levels (\autoref{fig:bike_hens_examples}-b). This time, the avatars were chosen randomly but in a way that matches the victim's gender. By walking among the victims in the ``hospitalized'' bar, the user can see that they were mostly men, which suggests a potential gender effect. 

Users can again immerse themselves in a 360° view of the place where each accident took place, which can help them build a deeper, more qualitative understanding of the circumstances of the accident. For example, in one case that resulted in a death, the infobox indicates that the accident occurred at night, without public lighting. By immersing themselves into the 360° view, the user can learn that the accident happened in a long road in the forest where cars are probably going fast. For the second case where the person was killed, the infobox indicates that the accident occurred during the day under a heavy rain. The 360° view reveals that it happened on a parking lot, a piece of information that could not be deduced from the structured data alone.

\subsection{Use Case 3: Hen Welfare}
\label{sec: animal use case}

To illustrate that ZEVs can be used beyond human individuals, we applied the concept to the topic of animal well-being. Here, a ZEV is used to visualize agregated data about laying-hen farming in France. We collected data on different websites, and for each farming type (five categories, from enriched cages to organic), we recorded basic information (legally-mandated indoor and outdoor densities, maximum herd size, and average egg price), as well as the percentage of farms in that category. We then created a CSV file for 100 hypothetical farms matching the proportions of the five farming types. \autoref{fig:bike_hens_examples} (c and d) shows screenshots.

This ZEV again starts with a bar chart, this time showing the distribution of farms by farming type. When the user zooms in, the bar chart turns into a unit chart where each square stands for one of the 100 hypothetical farms. The next transition differs from the previous use cases and is more complex (see \autoref{fig:bike_hens_examples}-c); When zooming in further, the squares disappear, the bars become textured (concrete floor when there is no outdoor access, and grass otherwise), and hen avatars appear on top of the textured bars, with their density matching the farming density (either indoor or outdoor, depending on outdoor access). Thus, there is a shift of unit of visualization (from individual farms to individual hens), and this time, the density of the 3D avatars encodes data.

In this use case, walking among the avatars helps the user get a better sense of hen density. The user can then grab a sphere to immerse themselves in a 360° photo of a typical hen farm in the given farming category. This last phase is meant to give the user a more visceral impression of the density and living conditions of hens and presumably facilitate perspective-taking.

\subsection{Design methodology}
We implemented a generic ZEV prototype that can be applied to various datasets without additional coding. A video explaining how to configure a new visualization is available in our \href{https://osf.io/awqgc/}{OSF repository}.

Importantly, our focus was on implementing a general-purpose ZEV prototype that can be applied to a range of different datasets, rather than developing finalized ZEV examples that are ready to be deployed and used to raise public awareness. Since our goal was to illustrate the general potential of ZEVs for research purposes, we applied our generic prototype to several existing datasets, without extensively tailoring or optimizing the design for each dataset. In particular, we did not involve experts and stakeholders on any of the topics we used to illustrate ZEVs. However, for actual ZEV applications that are meant to be broadly deployed, it is very important to involve stakeholders in the design, to ensure that the right decisions are made about which information to share, how to represent the victims, and which immersive scene to display.

\subsection{Implementation Details}
\label{sec:implementation}

The prototype was implemented with Unity 2022.3.1f1 using C\# and SteamVR\footnote{\url{https://store.steampowered.com/app/250820/SteamVR/}}. The prototype runs on an Oculus Quest 2 and a laptop computer MSI GT63 Titan 10SF with an Intel Processor Core i7-10750H CPU 2.59 GHz, 16Go RAM and a Nvidia GTX 2070 graphic card. We used avatar models from Mixamo\footnote{\url{https://www.mixamo.com/}}, Microsoft Rocketbox\footnote{\url{https://github.com/microsoft/Microsoft-Rocketbox}} and the Inria avatar gallery\footnote{\url{https://avatar.inria.fr/}}. All models were processed with Blender to create Levels Of Details so they can be used on Unity. All animations were downloaded from Mixamo. We developed a simple ZEV builder that allows authors to provide other datasets in the form of CSV files and configure the resulting visualizations using the Unity inspector. %
Our code is available as open source (see information in our \href{https://osf.io/awqgc/}{OSF repository}).

\section{ZEVs -- The General Concept}
\label{sec: concept}

We discussed three concrete examples of ZEVs. Yet, the prototype we implemented and used for our three use cases (\autoref{sec: use cases}) is one possible implementation of ZEV, and other designs could fit into the definition of a ZEV. Here we lay out the general concept, discuss what the previous examples have in common, and attempt to generalize beyond those examples. We will occasionally use concepts from Morais et al.'s \cite{morais2020showing} design space of anthropographics to support our discussions.

\subsection{General Definition}

We tentatively define Zoomable Empathic Visualizations (ZEVs) as \textit{\textbf{interactive experiences} that convey information about \textbf{groups of individuals}, and use \textbf{semantic zooming} to support seamless navigation between \textbf{different views}, from aggregated statistical visualizations to representations that emphasize individuals}. The latter tend to be more qualitative, relatable, concrete, and immersive. The intent of ZEVs is to promote a holistic understanding of the situation of a group of individuals, which combines a quantitative and rational understanding of the general trends with a contextualized and visceral understanding of the individual situations. We further elaborate on the concept in the next subsections.

\subsection{Groups of Individuals}

By our definition, ZEVs convey information about groups of individuals. A major goal of ZEVs is to facilitate perspective-taking and promote empathy and compassion, which implies a focus on sentient beings. Both humans and non-human animals (e.g., laying hens) qualify, but inanimate objects do not. Visualizations focusing on single individuals are also excluded, as the focus is on covering both the group level and the individual level.

Ideally, for a ZEV, a dataset must be provided that is not aggregated and contains information about each individual, as in our femicide and bicycle accidents examples. ZEVs can still be used when personal information is obfuscated for privacy purposes (as in our femicide example), or when the individuals are hypothetical or fictional, for example because only aggregated data is available (as in our laying hens example). 

In this article, we focus on communicating about negative topics because this is where we think ZEVs may be the most useful. However, nothing in our definition excludes neutral or positive topics---for example, ZEVs may be used to communicate about people who feel happy or fortunate because they got cured from a serious illness or received help from a fund.

\subsection{Semantic Zooming}

By our definition, ZEVs rely on semantic zooming to navigate between different views, which excludes other forms of navigation (e.g., slideshows), except as a complement. However, there are many ways semantic zooming can be implemented. 

Besides the VR-based interactions we described, one could use proxemic interaction \cite{vogel2005interactive,isenberg2013hybrid}, e.g., showing a statistical chart on a wall display and having individuals appear when the user approaches the display. In addition, the starting point does not need to be an aggregated visualization: an alternative could be to show an individual first, and then let users zoom out to see the entire group \cite{dragicevic2022towards}.

\subsection{Different Views}
\label{sec:different_views}

Any type of view may be included in a ZEV, although the principle is always to cover a continuum between aggregated representations and representations that emphasize individuals, in a way that blends structured data with unstructured qualitative information.

\subsubsection{Examples of Zoom Levels} \label{sec:exampleZoomLevels}

Our prototype uses five distinct zoom levels. We describe them and discuss possible variations:

\begin{enumerate}%
    \item A \textbf{regular bar chart} that shows how individuals are distributed according to a feature of interest, e.g., the age of a femicide victim, or the gravity of a bicycle accident. At this level, it is generally useful to allow users to explore the dataset by selecting other features. Any other standard visualization may be used; we focused our examples on bar charts for simplicity.
    
    \item A \textbf{symbolic unit chart} that represents individuals using simple symbols (squares in our examples). This second level adds new information, which is the number of individuals for each category, and possibly more if different symbols are used to encode data attributes.
    
    \item An \textbf{anthropomorphic unit chart} that represents individuals using realistic 3D avatars. This level is mostly an intermediate step preparing the user for level 4, as the avatars are distant and their details hard to see. We discuss different approaches for choosing 3D avatars later on.
    
    \item An \textbf{immersive unit chart} which is the same as level 3, except the user is situated among the avatars, as in the mass shooting visualization from Ivanov et al. \cite{Ivanov2019AWalk}. Now the user can zoom into individuals simply by walking towards them. They can look at them more closely and get more individual information. In our case, we used an infobox showing extra information with text, as well as photos, in an attempt to amplify the identifiable victim effect \cite{Lee2016Identifiable}. Interactive avatars, sound, and videos are possible too.
    
    \item An \textbf{immersive view} of a place relevant to the individual and the topic of the ZEV, e.g., the neighborhood where a femicide victim died. This view can provide further details and can encourage the user to take the perspective of the individual. It is also possible to provide an interactive immersive scene in which the user impersonates the victim. VR has been used in such a manner in a past study \cite{seinfeld2018offenders}, where males with a history of domestic violence were asked to virtually embody a female avatar, and to witness a virtual simulation of a scene of abuse from a first-person perspective.
\end{enumerate}

Our current implementation of ZEV uses the five zoom levels described above; however, depending on the topic and design, more or fewer levels could be implemented. For instance, the immersive view could be split into two levels, starting with a third-person point of view and transitioning to a first-person point of view where the user embodies the individual.

\subsubsection{Mapping Avatars}

The 3D avatars used in levels 3 and 4 can be selected from avatar libraries and mapped to data points by either: \emph{i)} assigning avatars randomly; \emph{ii)} assigning avatars based on data attributes (e.g., gender or age), \emph{iii)} assigning avatars based on their physical resemblance with the actual individuals.

\subsubsection{Structured data vs. Unstructured Qualitative Information}

Regardless of the views chosen, a recurring theme and key aspect of ZEVs is the use of unstructured qualitative information to complement structured data. A photo portrait, a physically-resembling 3D avatar, or a 360° photo carry rich visual information that can never be fully encoded in structured datasets. A similar philosophy can be found in \textit{Dollar Street} \cite{DollarStreet}, a Web app that lets users explore data about the incomes and living conditions of families around the world, but also contains photos and videos depicting the families' way of living. Besides its informativeness, unstructured qualitative information can sometimes be interpreted more directly and can carry more weight. For example, the artistic installation \textit{what were you wearing?} \cite{whatWereYouWearing} shows the clothes victims were wearing when they got sexually assaulted; The purpose is to \textit{``to challenge the idea that provocative clothing is the cause of the sexual assault''}, in a way that is much more direct and effective than by showing data. Unstructured qualitative information---and stories in particular---can also help re-humanize victims. For example, Sarah Barukh wrote a book where she asks 125 women personalities to write the story of a woman who died from a femicide \cite{barukh2023}, because she wanted to present victims as more than \textit{``a number of stab wounds''} \cite{bayt2024vivantes}. We could imagine that in future ZEVs, when the user approaches an avatar, it comes to life, narrates their story, and even interacts with the user.

\section{User Study}
\label{sec:study}

To gather feedback and insights on how people may use and experience this new type of visualization, we conducted an exploratory user study in which participants experienced our ZEV prototype covering femicide data. %
This study received ethical approval from the ethical committees of Inria (Avis COERLE n$^\circ$ 2024-13) and of the University of Bordeaux (CER-UB-2024-5A-F).

\subsection{Participants}

We recruited 12 French-speaking participants (6 female, 6 male, aged 22--43, mean = 30, sd = 7) through professional and personal networks. 
Ten out of the twelve participants reported having no or low experience in VR (3 \textit{None} and 7 \textit{Low}), and the remaining two participants had an average experience.

Because the ZEV explores a topic that can resonate differently with individuals depending on gender, we refer to participants as F or M—based on their self-identified gender—along with their participant number.
More information about the participants (age, gender, VR experience and domain of study or work) are available in the supplementary material.

\subsection{Procedure}
Although our goal was to gather feedback on the general concept of ZEV, we needed to select a specific topic. We chose femicides (\autoref{sec: use case feminicides} and \autoref{fig:teaser}) because we believed it to be the example where access to qualitative information about individuals would have the greatest educational and emotional impact.

Potential participants who were at risk of being strongly impacted by the topic covered were carefully screened out, including through a clear warning in the recruitment message. Once in the lab, participants signed a consent form and filled a short demographic questionnaire. They then went through a training session to familiarize themselves with the ZEV navigation techniques, after which they were invited to explore the femicide dataset. We chose not to anonymize the victims' names and photos, as fabricated identities are thought to elicit weaker emotional engagement \cite{morais2021can}. Participants were first guided through the zoom levels one by one, and were asked to answer questions about the data at each level. They were then free to explore the visualization on their own for as long as they wanted to.

This phase was followed by an audio-recorded semi-structured interview where the experimenter invited participants to share the insights they gained, the emotions they experienced, and their opinions regarding this type of intereactive visualization. After the interview, participants were debriefed and given the option to read a summary of a governmental report on violent deaths among couples in France in 2022 \cite{GovernmentalReport}, should they wish to learn more.

\subsection{Analysis Methodology}

We transcribed the audio interviews, and analyzed them using Braun and Clarke's thematic analysis method \cite{Braun2006thematic}. Following the method's steps: \emph{1)} we familiarized ourselves with the material by reading all transcriptions; \emph{2)}  the main author coded the transcriptions, by extracting 767 sections of interest which were associated with 78 codes using an iterative method; 
\emph{3)} we aggregated the codes into patterns, which yielded themes; \emph{4)} we reviewed those themes, and finally \emph{5)} we named and defined our final themes. The results are discussed next. All the excerpts are translated from French. 
Supplementary material, including the interview questions, thematic analysis material, and the excerpts are available in our \href{https://osf.io/awqgc/}{OSF repository}.

\subsection{Generating Insights}

\paragraph{Learning from the Statistical Charts}
Several participants already made discoveries during this first stage of the ZEV. For example, F9 \excerpt{[realized] that it can happen at any age}, and F10 was similarly \excerpt{surprised that there were so many femicides of older women}. For some participants, the immersive views further helped get a better sense of the numbers. For example, F11 mentioned that \excerpt{the crowd at the end makes you realize the extent of the number of femicides}

\paragraph{Learning from non-Structured Data}
Several participants reported learning from the 360° scenes. F7 mentioned that \excerpt{the 360 view really shows that it happens in very ordinary locations}. M3 initially thought that femicides mostly happened in big cities, and stated that with 360° views \excerpt{it was surprising that there were so many in the countryside}.

\subsection{Eliciting Empathy and Affect} \label{sec:elicitEmpathy}

\paragraph{Negative Emotions}
Most participants reported being emotionally affected by the experience. While the intensity varied, the emotions expressed were predominantly negative. We identified common themes in their responses, reflected in words and phrases related to \excerpt{horror and shock}, \excerpt{sadness}, \excerpt{dismay and incomprehension}, and \excerpt{anger}. A more thorough analysis of these themes is available in the supplementary material.

Several participants expressed a feeling of unfairness, often paired with anger. M2 mentioned \excerpt{all those wasted lives} and F9 said that \excerpt{people are exterminated for absurd reasons}. Two participants (M5, F11) also mentioned feeling helpless while exploring the visualization. Similarly, M5 mentioned \excerpt{I can't do anything about it} and F11 reported she \excerpt{felt like being a spectator and not being able to do anything to help}.

\paragraph{Empathy and Perspective-Taking}
As participants engaged with the avatars, many felt a deeper connection to the victims and the issue of femicide. For M12, with statistical charts people are \excerpt{out of sight, out of mind}, while seeing the avatars is a reminder of the real people behind the data. Similarly, M3 mentioned that even though the issue of femicide hasn't directly affected him or his social circle, the unit chart with the avatars made it feel more real, emphasizing that \excerpt{it is not a distant issue, we need to be aware of that}.

Several participants engaged in perspective-taking, in different ways. Some focused on similarities between the victims and themselves or their relatives, especially age and city of residence. For example, F1 mentioned \excerpt{she was 23, I thought it's my age}, while F7 remarked that \excerpt{the ages are really harrowing because there are people who are my age, so inevitably you put yourself in their shoes}. Other participants engaged in perspective-taking by imagining themselves as a first-person witness. Speaking of the 360° views, F11 mentioned \excerpt{we were where they got killed, it really felt like being part of the scene}. 

Surprisingly to us, several participants tried to take the perspective of the murderer as an attempt to make sense of the inconceivable. M3 stated: \excerpt{I did put myself a little bit into the killer's mind, to try to better understand} and further stated that for some causes of death, \excerpt{it means you really want to make her suffer}.

\subsection{Sense of Reality and Immersion}
\paragraph{Sense of Reality}
Some participants reported that the 3D avatars and the 360° view brought a sense of reality to the data. F1 stated that with the 3D avatars, \excerpt{it becomes more tangible when it is real people rather than cubes}, while M6 mentioned that the 360° view \excerpt{settles us even more into reality. We really are in a real place in France where it happened}, and reported that this stage was \excerpt{the most concrete}. 
By bringing a sense of reality, the zoomed-in ZEV humanizes the data. As M12 mentioned, \excerpt{it no longer is a statistic, it is an individual}. Several participants felt this was amplified when they were walking among the avatars. F7 shared \excerpt{I saw the avatars, I saw people. I was able to imagine the life they had}. As the avatars stand for deceased women, several participants drew analogies with cemeteries and memorials. M2 mentioned that \excerpt{you literally feel like being in a cemetery and you pay tribute to each person}. When exploring the 360° views, M12 compared it to \excerpt{a pilgrimage, you come to commemorate on the place}.

\paragraph{Effects of VR}
Several participants mentioned that VR made them forget about the external world. F10 commented it was \excerpt{like being submerged by the avatars, the visualization}. Similarly, M3 reported \excerpt{being isolated in your space. Every time I took off the headset I thought oh yes, right, I'm in this box}. VR also helped some participants focus by blocking external distractions. 
F7 explained \excerpt{we're really in immersion so it's great because you're focused on what you do, what you're looking at}. M4: \excerpt{You're not in front of a screen [...]. Here you're in the middle of people}.

Some participants also mentioned that VR can engage due to its novelty. For example, for M2, \excerpt{It could be a good experience because we think a VR headset, it's not common [...] and since it's surprising, people are curious and they could come}.

\subsection{Perceived Benefits of Semantic Zooming}

\paragraph{Multi-scale Exploration}
F10 mentioned that the statistical chart \excerpt{allows you to comprehend the information as a whole, and then you can easily target where you want to go}. For M12, the chart helped contextualize the individual representations, as \excerpt{you end up next to one, two or a few people. But you know there are more around. Since you zoomed before, you know a little, you're aware of how many people are around}.

\paragraph{Gradual Aspect}
Some participants mentioned appreciating the gradual aspect of the zooming interactions: 
F1 stated that she \excerpt{didn't directly go from a big graph to the avatars, [...] I was prepared progressively}. M4 also mentioned: \excerpt{First you prepared me, you gave me numbers like a rational man who reads an article. [...] and progressively you immersed me, and you talked to my emotions}.

\paragraph{Freedom of Exploration}
\label{sec:freedom_exploration}
Participants appreciated that the zooming interactions allowed them to dive deeper into parts of interest, while avoiding parts they were not prepared to see. F10 mentioned: \excerpt{We go directly where we want to. I mean, we see a piece of information we find interesting, we can directly target where to go and then move to look at something else, the information next to it}. F11 similarly mentioned: \excerpt{It felt like every time I wanted to get more details on something, I clicked and there was a way to see more details}. Meanwhile, for F8: \excerpt{You can look at what you want. If there is something that you don't want to see, you can choose not to go look at it. [...] You can keep your distance which allows you not to look at the photos, not to look at the age}, and she explained that it allows her to \excerpt{have various reading levels and get more or less involved}. M5 chose not to explore the 360° views as he was worried he could recognize some locations.

\subsection{Perceived Limitations of ZEVs}

\paragraph{Morbidity and Voyeurism}
Although no participant mentioned finding the ZEV inappropriate, some participants used words such as \excerpt{gloomy} (F1, M4, M5), \excerpt{macabre} (M12), \excerpt{morbid} (M3), \excerpt{intrusive} (F10), and \excerpt{voyeurism} (M4, M5) to describe their feelings during the experience. F10 did not explore 360° views other than the one we invited them to look at during the guided stage, because she \excerpt{felt it would be intrusive to look at where these women died}.

\paragraph{Limitations of VR}
M6 and M12 stated feeling a bit nauseous at the end of the experience, and F1 reported having a slight headache. Although those are common side-effects of VR, they can add up to the emotional fatigue caused by the content of the experience. 

Four participants (M2, M6, F10, M12) noted that a drawback of VR is its limited accessibility, e.g., F10: \excerpt{the visualization is interesting provided that you have the equipment and know how to use it, etc. And it is costly and therefore not accessible to all}. Thus, we believe that such visualizations could be used in public events to generate interest in the subject, and later, people could consult more standard sources of information at home.

\paragraph{Limitations of the Current Design} \label{sec:designlimits}
Some limitations were mentioned that were specific to our current ZEV design, rather than being inherent to the ZEV concept. For example, we chose not to include a $y$-axis in our first chart because we wanted to reveal the numbers in the unit chart, but some participants found it disturbing, e.g., M2: \excerpt{it wasn't really indicative because you didn't have the numbers}. 
 Finally, the limited realism and diversity of the 3D avatars in our proof-of-concept prototype negatively impacted the experience of some of our participants. We explained our avatar matching process to any participant who asked, and as a result, M3 lost interest in the avatars: \excerpt{knowing that the avatars weren't representative, it made me stop looking at them. If I had seen the real face, I think it would have been a lot more striking}.

\subsection{General Positive Reactions}

Despite the heavy nature of the topic and the negative emotions experienced by most participants, we received lots of positive feedback about the overall ZEV experience. M6, F10 and M12 explicitly mentioned being glad they had this experience, and five participants (F1, M3, F7, F9, M12) thought it was important or a good thing to create and provide this kind of interactive visualization.

\paragraph{A Tool for Raising Awareness}
F9 stated that the ZEV \excerpt{is a great way to raise awareness. It speaks for itself. It humanizes the acts} and F10 shared that \excerpt{if we want people to react, and if we want them to be fully aware of what's happening and of the extent of the issue, I think it's perfect because it shakes us to see that}. 
M2 and F7 thought that more people should experience the visualization.

\paragraph{The Right Amount of Shock} \label{sec:amountShock}
Regarding the content and choice of representation in the visualization, F1 felt like the visualization had the right balance in terms of shock for the participants and explained that \excerpt{we are shocked in the sense that we think that this person really died and it's horrible, but not shocked in the sense that we did not see any atrocious picture and are about to have nightmares about it}. M3 compared the visualization to war documentaries, \excerpt{it's a little bit like when you are shown pictures of a conflict or cities that got bombed for instance. It feels a little bit the same. You don't want to see it but maybe it is necessary.}

\subsection{Study Takeaways}

Our qualitative study confirmed our expectation that our ZEV prototype on the topic of femicides would create a strong impression on users. Most participants reported experiencing strong negative emotions, including participants who orally reported not being particularly sensitive to the topic of femicides. Despite this, no participant reported regretting having had this experience. For many, it was instructive and eye-opening, and some saw the ZEV as a powerful tool that could be used more broadly to raise public awareness on a difficult but important topic. Participants particularly appreciated the zooming metaphor, which they immediately understood, and they quickly saw how the different views complement each other. They gained insights from the abstract charts while also valuing the deeper information about individuals, which they found both instructive and more humane, as it promoted empathy and perspective-taking, while avoiding reducing victims to mere statistics. On the flip side, some participants felt that the detailed views had a somewhat macabre tone, with some describing them as potentially intrusive or voyeuristic. Yet, participants welcomed the freedom offered by the zooming metaphor, which allowed them to go deeper where they wanted but also avoid content they were not ready to see.

\section{Discussion}
We now discuss potential limitations and difficulties with ZEVs, especially for real applications, and perspectives for future work.

\subsection{Managing the Different Facets of Empathy}

We illustrated how ZEVs can facilitate empathy, but empathy is complex and can take different forms. Kaplan~\cite{Kaplan2008Global} identifies three possible responses to viewing tragedies: 1) vicarious trauma, \textit{``a response in which the viewer is shocked to the extent of being emotionally over-aroused''}; 2) empty empathy, which is \textit{``empathy that does not result in pro-social behavior''}, and 3) witnessing, \textit{``a response that transforms the viewer in a positive prosocial manner''}. Similarly, Bloom~\cite{bloom2017empathy} defines empathy as experiencing others' emotions without necessarily being drawn to prosocial behavior, and contrasts it with compassion, \textit{``the feeling that arises in witnessing another's suffering and that motivates a subsequent desire to help''}.

For ZEVs to be useful, they must promote witnessing and compassion while avoiding both indifference and negative effects of empathy, such as distress and avoidance. Designing ZEVs to achieve this is challenging. In our prototypes, we minimized the risk of vicarious trauma by avoiding graphic or violent content and carefully warned potential participants about the topic of the ZEV. Participants reported a range of empathy-related emotions, but our study was not designed to disentangle the contributions of vicarious trauma, empty empathy, and witnessing/compassion. However, one participant indicated an intention to share their experience with friends for prevention and awareness, suggesting prosocial behavior. Additionally, one participant noted that the ZEV produced an appropriate level of shock, while another appreciated that the controllable zoom factor allowed them to limit exposure to what they could emotionally handle.

\subsection{Ethical Issues in Designing ZEVs for Real-World Deployment}

The use cases presented in this paper were intended solely to illustrate and test the possibilities offered by the ZEV concept. However, when a ZEV is intended to be deployed for public use on sensitive topics such as femicides, important ethical considerations should be taken into account. Designers must carefully consider what data should be shown and how, ensuring that it respects the victims and their family, as well as the end users. In contrast to previous work specifically focusing on ethics \cite{de2004picture, Nilsson2020Ethics}, the goal of this article is not to offer a thorough treatment of ethical considerations; Nevertheless, we highlight three important ethical considerations for the design of ZEVs:

\begin{enumerate}
    \item For serious topics, the ZEV design process should involve all the relevant stakeholders. Representatives of the population represented should be involved in all of the design's steps, along with ethicists and regulatory bodies. Moreover, if the represented individuals are still alive, designers should collect their consent.

    \item As access to rich and reliable data about victims is often impossible, ZEV designers may be unable to guarantee the full veracity of immersive scenes, similar to documentaries and biographical films which reconstruct events from partial information and must “fill in the gaps”. Designers must ensure that any fabricated elements are clearly acknowledged and do not harm the represented populations.

    \item ZEV design should be carefully tailored to both the topic and the target audience. Topics differ in sensitivity, scope of affected populations, and availability of data, requiring systematic adaptation of the design process. Audience characteristics must also shape design choices. For example, as discussed in Section~\ref{sec:exampleZoomLevels}, letting participants embody victims---similar to Seinfeld et al.~\cite{seinfeld2018offenders}, where men convicted of domestic violence embodied a female avatar being verbally assaulted---may be ethically acceptable for potential criminals but not for potential victims.
\end{enumerate}

\subsection{Practical Challenges}

\paragraph{Collecting Detailed Information.} 
A major hurdle for ZEVs---and for reporting stories about individuals more broadly---is the challenge of collecting detailed information. Most datasets contain aggregated or anonymized data. For example, D’Ignazio et al. \cite{d2022feminicide} highlight how official institutions often neglect the collection of femicide data and emphasize the difficulties faced by independent organizations. Designing ZEVs that are both informative and impactful will require substantial information-gathering efforts, as exemplified by Sarah Barukh's book \textit{125 et des milliers} \cite{barukh2023}, the \textit{Dollar Street} web app \cite{DollarStreet}, or the art piece \textit{what were you wearing?} \cite{whatWereYouWearing}.

\paragraph{Achieving High Avatar Realism.} 
In our ZEV prototypes, the avatars exhibited a stiff pose and had very simple animations, which compromised their visual realism. This is a critical concern, as prior research has shown that technical limitations in representing victims can reduce audience empathy \cite{Schultz2023Creating}. Therefore, when developing a ZEV for real-world deployment, designers and developers should prioritize achieving a high level of realism in human avatars. However, whether avatars should resemble the actual victims is a difficult issue. In Section \ref{sec:designlimits}, we noted that one participant found the experience less engaging upon learning that the avatars’ appearances were not modeled after the victims. However, replicating victims’ likenesses raises significant ethical concerns.

\paragraph{Rendering Large Populations.}

A major implementation challenge we faced is the difficulty of rendering a large number of detailed avatars in Unity without exceeding the available processing power. Our prototype can display a few hundred avatars, while some datasets contain data about thousands of individuals \cite{missingmigrants}. As outlined previously, it is important that avatars are visually realistic, even in the presence of a very large number of avatars. One promising future direction is to rely on Unity's Data Oriented Technology Stack (DOTS) and the Entity Component System (ECS) \cite{DOTS}, which support advanced optimization techniques that should make it possible to render thousands of avatars in real-time.

\subsection{Need for More Studies}
Our ZEV study is only an initial qualitative evaluation. Here we discuss open research questions and directions for future studies.

\paragraph{Need for Better Evaluations.}

While our exploratory study contributes to the understanding of ZEVs, it is important that future work quantitatively assesses their merits and limitations. Such evaluations are particularly crucial given ongoing debates not only about the role of VR in fostering empathy \cite{christofi2022use}, but also about the effectiveness of unit charts \cite{morais2021can}. However, evaluating ZEVs presents challenges, including the wide range of possible baselines for comparison (e.g., conventional visualizations, paper reports, or video documentaries), the multiple views that may need to be evaluated independently (e.g., the unit chart, the crowd view, the 360° view), and the diversity of possible outcomes (e.g., knowledge acquisition and retention, affect, empathy and compassion, attitude change, and behavior).

\paragraph{How Critical is VR for Eliciting Empathic Responses?}

Several participants in our study noted that VR enhanced the sense of realism. Outside the study, colleagues who had seen a desktop version found the immersive experience notably more striking. However, further research is needed to systematically compare immersive ZEVs with their non-immersive counterparts in order to assess the specific benefits and drawbacks of immersion. Moreover, our study did not measure participants’ sense of presence---an important factor that should be considered in future investigations.

\section{Conclusion}

We introduced zoomable empathic visualizations (ZEVs), a new type of interactive experience allowing users to seamlessly navigate between abstract statistical visualizations and representations that are more qualitative, immersive, and focused on individuals. 
We presented three use cases of ZEVs involving communication about human and animal tragedies, and reported on a qualitative study that offers empirical insight into the potential benefits of visualizations that emphasize individuals. Results suggest
 that ZEVs can be powerful tools for educating and raising awareness on difficult issues.

\section*{Supplementary materials}
\label{sec:supMaterial}
Supplementary material for this article is available at: 

\href{https://osf.io/awqgc/}{https://osf.io/awqgc/}.

\bibliographystyle{unsrt}  
\bibliography{bibliography}

@book{slovic2015numbers,
  title={Numbers and nerves: Information, emotion, and meaning in a world of data},
  author={Slovic, Paul and Slovic, Scott},
  year={2015},
  publisher={Oregon State University Press}
}

@book{bucher2017storytelling,
  title={Storytelling for virtual reality: Methods and principles for crafting immersive narratives},
  author={Bucher, John},
  year={2017},
  publisher={Routledge}
}

@article{Hiltunen2019Documentary,
author = {Kaisa Hiltunen},
title = {Recent documentary films about migration: in search of common humanity},
journal = {Studies in Documentary Film},
volume = {13},
number = {2},
pages = {141-155},
year = {2019},
publisher = {Routledge},
doi = {10.1080/17503280.2019.1595919}
}

@article{small2003helping,
  title={Helping a victim or helping the victim: Altruism and identifiability},
  author={Small, Deborah A and Loewenstein, George},
  journal={Journal of Risk and uncertainty},
  volume={26},
  pages={5--16},
  year={2003},
  publisher={Springer},
doi={10.1023/A:1022299422219}
}

@misc{harris2015connecting,
  author = {Jacob Harris},
  title = {Connecting with the Dots},
  year = {2015},
  note = {Last accessed 2026-01-14},
  url = {https://source.opennews.org/articles/connecting-dots/}
}

@misc{slobin2014WhatIf,
  author = {Sarah Slobin},
  title = {Connecting with the Dots},
  year = {2014},
  note = {Last accessed 2026-01-14},
  url = {https://source.opennews.org/articles/what-if-data-visualization-actually-people/}
}

@article{vastfjall2014compassion,
  title={Compassion fade: Affect and charity are greatest for a single child in need},
  author={V{\"a}stfj{\"a}ll, Daniel and Slovic, Paul and Mayorga, Marcus and Peters, Ellen},
  journal={PloS one},
  volume={9},
  number={6},
  pages={e100115},
  year={2014},
  publisher={Public Library of Science San Francisco, USA},
doi={10.1371/journal.pone.0100115}
}

@article{kogut2005identified,
  title={The “identified victim” effect: An identified group, or just a single individual?},
  author={Kogut, Tehila and Ritov, Ilana},
  journal={Journal of behavioral decision making},
  volume={18},
  number={3},
  pages={157--167},
  year={2005},
  publisher={Wiley Online Library}
}

@article{bleiker2013visual,
  title={The visual dehumanisation of refugees},
  author={Bleiker, Roland and Campbell, David and Hutchison, Emma and Nicholson, Xzarina},
  journal={Australian journal of political science},
  volume={48},
  number={4},
  pages={398--416},
  year={2013},
  publisher={Taylor \& Francis},
  doi = {10.1080/10361146.2013.840769}
}

@article{slovic2007if,
  title={“If I look at the mass I will never act”: Psychic numbing and genocide},
  author={Slovic, Paul},
  journal={Judgment and Decision making},
  volume={2},
  number={2},
  pages={79--95},
  year={2007},
  publisher={Cambridge University Press},
doi={10.1017/S1930297500000061}
}

@article{arcagni2021vr,
  title={VR storytelling: Potentials and limitations of virtual reality narratives},
  author={Arcagni, Simone and D'Aloia, Adriano},
  journal={Cinergie--Il Cinema e le altre Arti},
  number={19},
  pages={1--7},
  year={2021}
}

@inproceedings{christofi2022use,
  author={Christofi, Maria and Hadjipanayi, Christos and Michael-Grigoriou, Despina},
  booktitle={IMET'22}, 
  title={The Use of Storytelling in Virtual Reality for Studying Empathy: A Review}, 
  year={2022},
  volume={},
  number={},
  pages={1-8},
  doi={10.1109/IMET54801.2022.9929546}
}

@misc{united2016clouds,
  author = {{United Nations, Within}},
  title = {Clouds Over Sidra},
  organization = {Youtube},
  year = {2016},
  note = {Last accessed 2026-01-14},
  howpublished = "\href{https://www.youtube.com/watch?v=mUosdCQsMkM}{Youtube link}"
}

@misc{united2019Haiti,
  author = {{United Nations, MINUJUSTH}},
  title = {Haiti My destiny as a kid at risk},
  organization = {Youtube},
  year = {2019},
  note = {Last accessed 2026-01-14},
  howpublished = "\href{https://www.youtube.com/watch?v=4ntggdu-l_I}{Youtube link}"
}

@misc{centerEResearch2022cut,
  author = {{Virtual Human Interaction Lab, Center for eResearch}},
  title = {1,000 Cut Journey},
  organization = {Youtube},
  year = {2022},
  note = {Last accessed 2026-01-14},
  howpublished = "\href{https://www.youtube.com/watch?v=UMERGKMw0FE}{Youtube link}"
}

@article{morais2020showing,
  title={Showing data about people: A design space of anthropographics},
  author={Morais, Luiz and Jansen, Yvonne and Andrade, Nazareno and Dragicevic, Pierre},
  journal={IEEE TVCG},
  volume={28},
  number={3},
  pages={1661--1679},
  year={2020},
  publisher={IEEE},
doi={10.1109/TVCG.2020.3023013}
}

@article{dragicevic2022towards,
  title={Towards immersive humanitarian visualizations},
  author={Dragicevic, Pierre},
  journal={arXiv preprint arXiv:2204.01313},
  year={2022}
}

@article{lee2020data,
  title={Data visceralization: Enabling deeper understanding of data using virtual reality},
  author={Lee, Benjamin and Brown, Dave and Lee, Bongshin and Hurter, Christophe and Drucker, Steven and Dwyer, Tim},
  journal={IEEE TVCG},
  volume={27},
  number={2},
  pages={1095--1105},
  year={2020},
  publisher={IEEE},
doi={10.1109/TVCG.2020.3030435}
}

@article{Braun2006thematic,
author = { Virginia   Braun  and  Victoria   Clarke },
title = {Using thematic analysis in psychology},
journal = {Qualitative Research in Psychology},
volume = {3},
number = {2},
pages = {77-101},
year  = {2006},
publisher = {Routledge},
doi = {10.1191/1478088706qp063oa},
URL = { 
        https://www.tandfonline.com/doi/abs/10.1191/1478088706qp063oa
},
eprint = { 
        https://www.tandfonline.com/doi/pdf/10.1191/1478088706qp063oa
}
}

@article{dragga2001cruel,
  title={Cruel pies: The inhumanity of technical illustrations},
  author={Dragga, Sam and Voss, Dan},
  journal={Technical communication},
  volume={48},
  number={3},
  pages={265--274},
  year={2001},
  publisher={Society for Technical Communication}
}

@misc{whatWereYouWearing,
  author={Brockman, Jen and Wyandt-Hiebert, Mary},
  title = {What Were You Wearing?},
  howpublished = {\href{https://sbaproject.org/what-were-you-wearing/}},
  note = {Accessed: 2026-01-14}
}

@misc{DollarStreet,
    author = "Anna Rosling-Rönnlund",
    title = "Dollar Street",
    howpublished = {\href{https://www.gapminder.org/dollar-street}{Visualization link}},
    note = {Accessed: 2026-01-14}
  }

@article{seinfeld2018offenders,
  title={Offenders become the victim in virtual reality: impact of changing perspective in domestic violence},
  author={Seinfeld, Sofia and Arroyo-Palacios, Jorge and Iruretagoyena, Guillermo and Hortensius, Ruud and Zapata, Luis E and Borland, David and de Gelder, Beatrice and Slater, Mel and Sanchez-Vives, Maria V},
  journal={Scientific reports},
  volume={8},
  number={1},
  pages={2692},
  year={2018},
  publisher={Nature Publishing Group UK London},
  doi={10.1038/s41598-018-19987-7}
}

@article{Lee2016Identifiable,
author = {Seyoung Lee and Thomas Hugh Feeley},
title = {The identifiable victim effect: a meta-analytic review},
journal = {Social Influence},
volume = {11},
number = {3},
pages = {199-215},
year = {2016},
publisher = {Routledge},
doi = {10.1080/15534510.2016.1216891},
URL = {https://doi.org/10.1080/15534510.2016.1216891},
eprint = {https://doi.org/10.1080/15534510.2016.1216891}
}

@misc{ Femicides,
    author = "{Feminicide par compagnon ou ex}",
    title = {"Les feminicides par compagnon ou ex publies en 2022 "},
    howpublished = "\href{https://www.feminicides.fr/}{Link}",
    note = "[Last accessed 2026-01-14]"
  }

@misc{BikeAccidents,
    author = "{Koumoul}",
    title = {"Accidents de vélo"},
    howpublished = "\href{https://www.data.gouv.fr/fr/datasets/accidents-de-velo/\#/resources}{Link}",
    note = "[Last accessed 2026-01-14]"
  }

@inproceedings{Boy2017Showing,
author = {Boy, Jeremy and Pandey, Anshul Vikram and Emerson, John and Satterthwaite, Margaret and Nov, Oded and Bertini, Enrico},
title = {Showing People Behind Data: Does Anthropomorphizing Visualizations Elicit More Empathy for Human Rights Data?},
year = {2017},
isbn = {9781450346559},
publisher = {ACM},
address = {New York, NY, USA},
url = {https://doi.org/10.1145/3025453.3025512},
doi = {10.1145/3025453.3025512},
booktitle = {Proceedings of CHI '17},
pages = {5462–5474},
numpages = {13},
location = {Denver, Colorado, USA}
}

@inproceedings{Dhawka2023WeAreTheData,
author = {Dhawka, Priya and He, Helen Ai and Willett, Wesley},
title = {We are the Data: Challenges and Opportunities for Creating Demographically Diverse Anthropographics},
year = {2023},
isbn = {9781450394215},
publisher = {ACM},
address = {New York, NY, USA},
url = {https://doi.org/10.1145/3544548.3581086},
doi = {10.1145/3544548.3581086},
booktitle = {Proceedings of CHI '23},
articleno = {807},
numpages = {14},
location = {, Hamburg, Germany, }
}

@ARTICLE{Ivanov2019AWalk,
  author={Ivanov, Alexander and Danyluk, Kurtis and Jacob, Christian and Willett, Wesley},
  journal={IEEE Computer Graphics and Applications}, 
  title={A Walk Among the Data}, 
  year={2019},
  volume={39},
  number={3},
  pages={19-28},
  doi={10.1109/MCG.2019.2898941}}

@book{barukh2023,
  title     = "125 et des milliers",
  author    = "Barukh, Sarah",
  year      = 2023,
  publisher = "Haper Collins",
  address   = "Paris"
}

@article{d2022feminicide,
  title={Feminicide and counterdata production: Activist efforts to monitor and challenge gender-related violence},
  author={D'Ignazio, Catherine and Crux{\^e}n, Isadora and Val, Helena Su{\'a}rez and Cuba, Angeles Martinez and Garc{\'\i}a-Montes, Mariel and Fumega, Silvana and Suresh, Harini and So, Wonyoung},
  journal={Patterns},
  volume={3},
  number={7},
  year={2022},
  publisher={Elsevier},
doi = {https://doi.org/10.1016/j.patter.2022.100530}
}

@misc{GovernmentalReport,
    author = {{French Ministry of Interior and Overseas Territories}},
    title = {"Étude nationale sur les morts violentes au sein du couple pour l'année 2022 "},
    howpublished = "\url{https://mobile.interieur.gouv.fr/Publications/Securite-interieure/Etude-nationale-sur-les-morts-violentes-au-sein-du-couple-pour-l-annee-2022}",
    note = "[Last; accessed 2026-01-14]"
  }

@inproceedings{ens2021grand,
  author = {Ens, Barrett and Bach, Benjamin and Cordeil, Maxime and Engelke, Ulrich and Serrano, Marcos and Willett, Wesley and Prouzeau, Arnaud and Anthes, Christoph and B\"{u}schel, Wolfgang and Dunne, Cody and Dwyer, Tim and Grubert, Jens and Haga, Jason H. and Kirshenbaum, Nurit and Kobayashi, Dylan and Lin, Tica and Olaosebikan, Monsurat and Pointecker, Fabian and Saffo, David and Saquib, Nazmus and Schmalstieg, Dieter and Szafir, Danielle Albers and Whitlock, Matt and Yang, Yalong},
title = {Grand Challenges in Immersive Analytics},
year = {2021},
isbn = {9781450380966},
publisher = {ACM},
address = {New York, NY, USA},
url = {https://doi.org/10.1145/3411764.3446866},
doi = {10.1145/3411764.3446866},
booktitle = {Proceedings of CHI'21},
articleno = {459},
numpages = {17},
location = {Yokohama, Japan}
}

@article{dibenigno2021flow,
  title={Flow immersive: A multiuser, multidimensional, multiplatform interactive COVID-19 data visualization tool},
  author={DiBenigno, Michael and Kosa, Mehmet and Johnson-Glenberg, Mina C},
  journal={Frontiers in psychology},
  volume={12},
  pages={661613},
  year={2021},
  publisher={Frontiers},
  doi={10.3389/fpsyg.2021.661613}
}

@article{Pena2010immersive,
  title={Immersive journalism: Immersive virtual reality for the first-person experience of news},
  author={De la Pe{\~n}a, Nonny and Weil, Peggy and Llobera, Joan and Spanlang, Bernhard and Friedman, Doron and Sanchez-Vives, Maria V and Slater, Mel},
  journal={Presence},
  volume={19},
  number={4},
  pages={291--301},
  year={2010},
  publisher={MIT Press},
  doi={10.1162/PRES\_a\_00005}
}

@inproceedings{weaver2004building,
  title={Building highly-coordinated visualizations in improvise},
  author={Weaver, Chris},
  booktitle={IEEE VIS'04},
  pages={159--166},
  year={2004},
  organization={IEEE},
doi={10.1109/INFVIS.2004.12}
}

@inproceedings{bederson1994pad++,
  author = {Bederson, Benjamin B. and Hollan, James D.},
title = {Pad++: a zooming graphical interface for exploring alternate interface physics},
year = {1994},
isbn = {0897916573},
publisher = {ACM},
address = {New York, NY, USA},
url = {https://doi.org/10.1145/192426.192435},
doi = {10.1145/192426.192435},
booktitle = {Proceedings of UIST '94},
pages = {17–26},
numpages = {10},
location = {Marina del Rey, California, USA}
}

@article{dunsmuir2009selective,
  author = {Dunsmuir, Dustin},
year = {2011},
month = {03},
pages = {},
title = {Selective Semantic Zoom of a Document Collection},
journal = {Simon Fraser University Master of Science Thesis}
}

@inproceedings{conti2005visual,
  title={Visual exploration of malicious network objects using semantic zoom, interactive encoding and dynamic queries},
  author={Conti, Gregory and Grizzard, Julian and Ahamad, Mustaque and Owen, Henry},
  booktitle={IEEE Workshop on VizSEC'05},
  pages={83--90},
  year={2005},
  organization={IEEE},
doi={10.1109/VIZSEC.2005.1532069}
}

@inproceedings{garcia2011semantic,
  title={Semantic zoom: A details on demand visualisation technique for modelling owl ontologies},
  author={Garcia, Juan and Theron, Roberto and Garcia, Francisco},
  booktitle={Highlights in Practical Applications of Agents and Multiagent Systems: 9th International Conference on Practical Applications of Agents and Multiagent Systems},
  pages={85--92},
  year={2011},
  organization={Springer}
}

@book{munzner2014visualization,
  title={Visualization analysis and design},
  author={Munzner, Tamara},
  year={2014},
  publisher={CRC press}
}

@article{rodriguez2020use,
  title={Use of virtual reality and 360° video as narrative resources in the documentary genre: Towards a new immersive social documentary?},
  author={Rodr{\'\i}guez-Fidalgo, Mar{\'\i}a Isabel and Pa{\'\i}no-Ambrosio, Adriana},
  journal={Catalan Journal of Communication \& Cultural Studies},
  volume={12},
  number={2},
  pages={239--253},
  year={2020},
  publisher={Intellect}
}

@inproceedings{verma2023designing,
  title={Designing Resource Allocation Tools to Promote Fair Allocation: Do Visualization and Information Framing Matter?},
  author={Verma, Arnav and Morais, Luiz and Dragicevic, Pierre and Chevalier, Fanny},
  booktitle={Proceedings of CHI '23},
  pages={1--16},
  year={2023}
}

@article{van2022news,
  title={Do news media kill? How a biased news reality can overshadow real societal risks, the case of aviation and road traffic accidents},
  author={van der Meer, Toni GLA and Kroon, Anne C and Vliegenthart, Rens},
  journal={Social forces},
  volume={101},
  number={1},
  pages={506--530},
  year={2022},
  publisher={Oxford University Press}
}

@inproceedings{morais2021can,
  title={Can anthropographics promote prosociality? A review and large-sample study},
  author={Morais, Luiz and Jansen, Yvonne and Andrade, Nazareno and Dragicevic, Pierre},
  booktitle={Proceedings of CHI '21},
  pages={1--18},
  year={2021},
doi = {10.1145/3411764.3445637}
}

@article{segel2010narrative,
  title={Narrative visualization: Telling stories with data},
  author={Segel, Edward and Heer, Jeffrey},
  journal={IEEE TVCG},
  volume={16},
  number={6},
  pages={1139--1148},
  year={2010},
  publisher={IEEE},
doi={10.1109/TVCG.2010.179}
}

@article{tong2018storytelling,
  title={Storytelling and visualization: An extended survey},
  author={Tong, Chao and Roberts, Richard and Borgo, Rita and Walton, Sean and Laramee, Robert S and Wegba, Kodzo and Lu, Aidong and Wang, Yun and Qu, Huamin and Luo, Qiong and others},
  journal={Information},
  volume={9},
  number={3},
  pages={65},
  year={2018},
  publisher={MDPI},
doi={10.3390/info9030065}
}

@inproceedings{perlin1993pad,
  author = {Perlin, Ken and Fox, David},
title = {Pad: an alternative approach to the computer interface},
year = {1993},
isbn = {0897916018},
publisher = {ACM},
address = {New York, NY, USA},
url = {https://doi.org/10.1145/166117.166125},
doi = {10.1145/166117.166125},
booktitle = {Proceedings of the 20th Annual Conference on Computer Graphics and Interactive Techniques},
pages = {57–64},
numpages = {8},
location = {Anaheim, CA},
series = {SIGGRAPH '93}
}

@misc{humdata2024,
  author = {The Centre for Humanitarian Data},
  title = {Center for Humanitarian Data -- Glossary},
  year = {2024},
  note = {Last accessed 2026-01-14},
  howpublished = {\href{https://centre.humdata.org/glossary/}{Link}}
}

@misc{stirlig2014,
  author = {Martin Stirling},
  title = {Most Shocking Second a Day Video (short movie)},
  year = {2014},
  note = {Last accessed 20 Jun 2024},
  howpublished = {\url{https://www.imdb.com/title/tt5009342/}, video available at \href{https://www.youtube.com/watch?v=fSIpARmq2WI}{Youtube link}.}
}

@misc{titanicVR2014,
  author = {Immersive VR Education Ltd},
  title = {Titanic VR (software)},
  year = {2018},
  note = {Last accessed 2026-01-14},
  howpublished = {\href{https://store.steampowered.com/app/741430/Titanic_VR/}{Steam link}, video available at \href{https://www.youtube.com/watch?v=Z4QV22VGfGs}{Youtube link}.}
}

@article{kilteni2012sense,
  title={The sense of embodiment in virtual reality},
  author={Kilteni, Konstantina and Groten, Raphaela and Slater, Mel},
  journal={Presence: Teleoperators and Virtual Environments},
  volume={21},
  number={4},
  pages={373--387},
  year={2012},
  publisher={MIT Press One Rogers Street, Cambridge, MA 02142-1209, USA journals-info~…}
}

@article{caviola2021psychology,
  title={The psychology of (in) effective altruism},
  author={Caviola, Lucius and Schubert, Stefan and Greene, Joshua D},
  journal={Trends in Cognitive Sciences},
  volume={25},
  number={7},
  pages={596--607},
  year={2021},
  publisher={Elsevier},
  doi = {https://doi.org/10.1016/j.tics.2021.03.015},
}

@inproceedings{bederson2011promise,
author = {Bederson, Benjamin B.},
title = {The promise of zoomable user interfaces},
year = {2010},
isbn = {9781450304368},
publisher = {ACM},
address = {New York, NY, USA},
url = {https://doi.org/10.1145/1865841.1865844},
doi = {10.1145/1865841.1865844},
booktitle = {Proceedings of VINCI '10},
articleno = {2},
numpages = {1},
location = {Beijing, China}
}

@article{assor2024augmented,
author = {Assor, Ambre and Prouzeau, Arnaud and Dragicevic, Pierre and Hachet, Martin},
title = {Augmented Reality Waste Accumulation Visualizations},
year = {2024},
issue_date = {June 2024},
publisher = {ACM},
address = {New York, NY, USA},
volume = {2},
number = {2},
url = {https://doi.org/10.1145/3636970},
doi = {10.1145/3636970},
journal = {ACM J. Comput. Sustain. Soc.},
month = may,
articleno = {11},
numpages = {29}
}

@article{park2017atom,
  title={Atom: A grammar for unit visualizations},
  author={Park, Deokgun and Drucker, Steven M and Fernandez, Roland and Elmqvist, Niklas},
  journal={IEEE TVCG},
  volume={24},
  number={12},
  pages={3032--3043},
  year={2017},
  publisher={IEEE},
doi={10.1109/TVCG.2017.2785807}
}

@phdthesis{vogel2005interactive,
  title={Interactive public ambient displays},
  author={Vogel, Daniel John},
  year={2005},
  school={University of Toronto}
}

@article{isenberg2013hybrid,
  title={Hybrid-image visualization for large viewing environments.},
  author={Isenberg, Petra and Dragicevic, Pierre and Willett, Wesley and Bezerianos, Anastasia and Fekete, Jean-Daniel},
  journal={TVCG},
  volume={19},
  number={12},
  pages={2346--2355},
  year={2013},
doi={10.1109/TVCG.2013.163}
}

@misc{FrenchDatasetInfant,
  author = {{Our World in Data}},
  title = {{Child mortality, 1800 to 2016, for France}},
  howpublished = {\href{https://ourworldindata.org/grapher/child-mortality?time=earliest..latest&country=~FRA}{Link}},
  note = {Last accessed 2026-01-14}
}

@article{owid-child-mortality-global-overview,
    author = {Hannah Ritchie},
    title = {How child mortality has declined in the last two centuries},
    journal = {Our World in Data},
    year = {2018},
    note = {https://ourworldindata.org/child-mortality-global-overview}
}

@misc{bayt2024vivantes,
  author = {Célyne Baÿt-Darcourt},
  title = {"Vivante(s)" sur Canal + : un mode d’emploi pour les victimes de violences conjugales},
  year = {2015},
  note = {Radio interview. Last accessed 9 September 2024},
  url = {https://www.francetvinfo.fr/replay-radio/info-medias/vivante-s-sur-canal-un-mode-d-emploi-pour-les-victimes-de-violences-conjugales_6376186.html}
}

@misc{missingmigrants,
  author = {{International Organization for Migration}},
  title = {{Missing Migrants Project}},
  year = {2024},
  note = {Last accessed 10 September 2024},
  url = {https://missingmigrants.iom.int/}
}

@misc{DOTS,
  author = {{Unity Technologies}},
  title = {Unity's Data-Oriented Technology Stack ({DOTS})},
  year = {2024},
  note = {Last accessed 10 September 2024},
  url = {https://unity.com/dots}
}

@misc{GoogleEarthVR,
  author = {{Google LLC}},
  title = {{Google Earth VR}},
  year = {2024},
  note = {Last accessed 10 September 2024},
  url = {https://store.steampowered.com/app/348250/Google_Earth_VR/}
}

@inproceedings{liem2020structure,
  title={Structure and empathy in visual data storytelling: Evaluating their influence on attitude},
  author={Liem, Johannes and Perin, Charles and Wood, Jo},
  booktitle={Computer Graphics Forum},
  volume={39},
  number={3},
  pages={277--289},
  year={2020},
doi = {https://doi.org/10.1111/cgf.13980},
  organization={Wiley Online Library}
}

@article{Kaplan2008Global,
author = {E. Ann Kaplan},
title = {Global trauma and public feelings: Viewing images of catastrophe},
journal = {Consumption Markets \& Culture},
volume = {11},
number = {1},
pages = {3--24},
year = {2008},
publisher = {Routledge},
doi = {10.1080/10253860701799918},
URL = { 
        https://doi.org/10.1080/10253860701799918
},
eprint = { 
        https://doi.org/10.1080/10253860701799918

}}

@article{de2004picture,
title={Picture perfect (?): Ethical considerations in visual representation},
  author={de Laat, Sonya},
  journal={NEXUS: The Canadian Student Journal of Anthropology},
  volume={17},
  number={1},
  year={2004},
  doi={10.15173/nexus.v17i1.192}
}

@article{Nilsson2020Ethics,
author = {Maria Nilsson},
title = {An Ethics of (not) Showing: Citizen Witnessing, Journalism and Visualizations of a Terror Attack},
journal = {Journalism Practice},
volume = {14},
number = {3},
pages = {259--276},
year = {2020},
publisher = {Routledge},
doi = {10.1080/17512786.2019.1623708},
URL = { https://doi.org/10.1080/17512786.2019.1623708},
eprint = { https://doi.org/10.1080/17512786.2019.1623708}
}

@article{Schultz2023Creating,
author = {Corey Kai Nelson Schultz},
title = {Creating the ‘virtual’ witness: the limits of empathy},
journal = {Museum Management and Curatorship},
volume = {38},
number = {1},
pages = {2--17},
year = {2023},
publisher = {Routledge},
doi = {10.1080/09647775.2021.1954980},
URL = { https://doi.org/10.1080/09647775.2021.1954980},
eprint = {https://doi.org/10.1080/09647775.2021.1954980}
}

@inproceedings{Mencarini2025Stories,
author = {Mencarini, Eleonora and Leonardi, Chiara and Massa, Paolo and D'Errico, Francesca},
title = {Stories from the Peaks: An Interactive Data Storytelling to Narrate Climate Change Impacts through a Pluralism of Voices},
year = {2025},
isbn = {9798400721021},
publisher = {Association for Computing Machinery},
address = {New York, NY, USA},
url = {https://doi.org/10.1145/3750069.3750130},
doi = {10.1145/3750069.3750130},
booktitle = {Proceedings of the 16th Biannual Conference of the Italian SIGCHI Chapter},
articleno = {11},
numpages = {8},
location = {
},
series = {CHItaly '25}
}

@article{bloom2017empathy,
  title={Empathy and its discontents},
  author={Bloom, Paul},
  journal={Trends in cognitive sciences},
  volume={21},
  number={1},
  pages={24--31},
  year={2017},
  publisher={Elsevier}
}

\end{document}